\documentclass[a4paper,11pt]{article}
\pdfoutput=1 
\usepackage[utf8]{inputenc}
\usepackage{geometry}
\usepackage[sumlimits,intlimits,namelimits]{amsmath}
\usepackage{amssymb}
\usepackage{a4wide}
\usepackage[english]{babel}
\usepackage{longtable, multirow,tabularx}
\usepackage[titletoc,title]{appendix}
\usepackage{slashed}
\usepackage{amsmath}
\usepackage{cite}

\makeatletter
\def\fmslash{\@ifnextchar[{\fmsl@sh}{\fmsl@sh[0mu]}}
\def\fmsl@sh[#1]#2{%
  \mathchoice
    {\@fmsl@sh\displaystyle{#1}{#2}}%
    {\@fmsl@sh\textstyle{#1}{#2}}%
    {\@fmsl@sh\scriptstyle{#1}{#2}}%
    {\@fmsl@sh\scriptscriptstyle{#1}{#2}}}
\def\@fmsl@sh#1#2#3{\m@th\ooalign{$\hfil#1\mkern#2/\hfil$\crcr$#1#3$}}
\makeatother
\usepackage{graphicx}
\usepackage{subfigure}
\usepackage{epstopdf}
\usepackage{xcolor}

\begin{document}
\begin{titlepage}
\begin{flushright}
SI-HEP-2020-09 \\[0.2cm]
SFB-257-P3H-20-015
\end{flushright}

\vspace{1.2cm}
\begin{center}
{\Large\bf
Heavy Quark Expansion for Heavy Hadron Lifetimes: \\[2mm]
Completing the \boldmath $1/m_b^3$ \unboldmath Corrections}
\end{center}

\vspace{0.5cm}
\begin{center}
{\sc Th.~Mannel, D. Moreno, and A.~A.~Pivovarov} \\[2mm]
{Theoretische Physik 1, Naturwiss. techn. Fakult\"at,
Universit\"at Siegen\\ D-57068 Siegen, Germany}  
\end{center}

\vspace{0.8cm}
\begin{abstract}
\vspace{0.2cm}\noindent
We complete the calculation of the contributions from the dimension six operators 
in the heavy quark expansion for the total lifetime of heavy hadrons. We give the leading order 
expressions for the Wilson coefficients of the Darwin term $\rho_D$ and the 
spin-orbit term $\rho_{\rm LS}$.
\end{abstract}

\end{titlepage}

\newpage
\pagenumbering{arabic}
\section{Introduction}
Inclusive weak decays of hadrons with a single heavy quark $Q$ have been intensively studied 
over the last decades~\cite{Lenz,Cheng:2018rkz,Krinner:2013cja}. 
The most inclusive quantity is the lifetime of the ground state 
heavy hadrons which is determined by their weak 
decay~\cite{Bagan:1994zd,Beneke:1998sy,Beneke:2002rj}. 
The theoretical method is the  
heavy quark expansion (HQE)~\cite{MW:00,G:04,N:94}, 
which is a combined expansion in the strong coupling 
$\alpha_s (m_Q)$~\cite{BG:95}
and inverse powers of the heavy quark mass~\cite{EH:90}. 
The leading term of 
this expansion is free of any hadronic parameter and is given by the decay rate of 
the ``free'' heavy quark. The corrections to this statement appear only at order 
$1/m_Q^2$ and are given in terms of the residual kinetic energy $\mu_\pi^2$ and 
the chromomagnetic moment $\mu_G^2$ which are both of order $\Lambda_{\rm QCD}^2$.  

Consequently these corrections should be at the level of a few percent, since the 
leading order result implies that all lifetimes of hadrons  with a single heavy 
quark $Q$ should be identical up to corrections of order $\Lambda_{\rm QCD}^2 / m_Q^2$. 
In the early days of the HQE this was taken as an embarrassment, since the lifetimes 
of the bottom hadrons had not been measured yet, and the lifetimes between charmed hadrons
differ by factors of two to five, which is related to the fact that the $c$-quark is too light for the HQE to be
a good approximation for these observables \cite{Fael:2019umf}. 

Since then the methods have been refined and the HQE makes quite precise predictions 
for the lifetime pattern of bottom hadrons and qualitatively describes the pattern of 
charmed hadrons. In fact, assuming $SU(2)_{\rm flavour}$ symmetry for the light quarks 
the lifetime differences between the three ground state mesons are driven by the terms 
of $1/m_Q^3$ and higher, in particular by the four-quark 
operators appearing at tree 
level, which involve light quarks of a particular flavour. 

The progress in the HQE for lifetimes rests on two pillars. On the one hand, there are
refinements in the HQE by including higher order terms in 
the $1/m_Q$ expansion~\cite{Heinonen:2014dxa}, on the
other hand there are perturbative calculations improving the Wilson coefficients appearing 
in the HQE. The higher orders 
in the HQE contain hadronic matrix elements, for which 
precise lattice predictions became available recently. Based on this, we are entering 
the precision era for these observables, in particular for the lifetime differences.   

However, a few ingredients have not yet been worked out in detail, since they were 
believed to be irrelevant. 
For this reason, the full calculation of all terms appearing at $1/m_Q^3$ has not yet 
been done, not even at tree level, since it was assumed that such terms will be small 
and mainly independent of the light-quark flavour. In the present paper we complete the
tree/level calculation of the $1/m_Q^3$ terms for the lifetime of a heavy hadron with 
a single heavy quark $Q$. While these contributions are known since some time for the 
inclusive semi-leptonic case, the full calculation of the terms at order $1/m_Q^3$ for 
the non-leptonic width was still missing.  

In section~\ref{Sec:synop} we describe the current status of bottom-hadron lifetimes.
In section~\ref{Sec:Outline} we give a short description of the method of the calculation         
of the non-leptonic width. In section~\ref{sect:results} we present our results, and discuss their 
implications in section~\ref{sect:discussion}.

\section{Synopsis on the status of bottom-hadron lifetimes}
\label{Sec:synop}

The measurements of lifetimes and lifetime rations for bottom hadrons have become 
very precise of the last decades. The current (2019) experimental averages 
obtained by the Heavy Flavor Averaging Group (HFLAV) of the $b$-hadron 
lifetime ratios are~\cite{Amhis:HFLAV}
\begin{equation}
 \frac{\tau(B_s)}{\tau(B_d)}\bigg|^{{\mbox{{\footnotesize exp}}}} = 0.994 \pm 0.004\,,
 \;\;\;\; 
 \frac{\tau(B^+)}{\tau(B_d)}\bigg|^{{\mbox{{\footnotesize exp}}}} = 1.076 \pm 0.004\,,
 \;\;\;\; 
 \frac{\tau(\Lambda_b)}{\tau(B_d)}\bigg|^{{\mbox{{\footnotesize exp}}}} = 0.969 \pm 0.006\,,
\end{equation}
which show that the experimental precision is indeed extremely high. 
Even higher precision seems to be 
achievable from the most recent results from LHCb~\cite{lhcb} and ATLAS~\cite{atlas}. 

The theory precision should of course live up to these experimental advancements.
The current status of the theoretical predictions is~\cite{Neubert:1996we,Lenz:2014,Lenz:2017}:
\begin{equation}
 \frac{\tau(B_s)}{\tau(B_d)}\bigg|^{{\mbox{{\footnotesize th}}}} = 1.0006 \pm 0.0025\,,
 \;\;\;\; 
 \frac{\tau(B^+)}{\tau(B_d)}\bigg|^{{\mbox{{\footnotesize th}}}} = 1.082^{+0.022}_{-0.026}\,,
 \;\;\;\; 
 \frac{\tau(\Lambda_b)}{\tau(B_d)}\bigg|^{{\mbox{{\footnotesize th}}}}
= 0.935 \pm 0.054\,,
\end{equation}
which shows that the HQE technique can be successfully applied to bottom hadron decays, 
allowing us to make precision predictions. 
Therefore, $B$-physics is entering in its precision era. To arrive at such precise
theoretical values, several advancements have been made: 

The leading term in the total decay rate, i.e. with the absence of power corrections, which describes the free $b$-quark decay and does not
contain non-perturbative corrections, is currently known at NLO-QCD~\cite{Hokim:84,Altarelli:91,Voloshin:95,Bagan:94,Bagan:95,LenzNierste:97,LenzNierste:99,LenzRauh:13} and at NNLO-QCD
in the massless case~\cite{Czarnecki:06} for non-leptonic decays. For semi-leptonic decays the current precision is 
NNLO-QCD~\cite{CzaneckiMelnikov:97,CzaneckiMelnikov:99,vanRitbergen:99,Melnikov:08,PakCzarmnecki:08,PakCzarmnecki2:08,Dowling:08,Bonciani:08,BiswasMelnikov:10,Caola:13}.

The contribution from the first power correction due to the dimension five kinetic and chromomagnetic operators is already known at LO-QCD 
for both semi-leptonic and non-leptonic decays~\cite{Uraltsev:92,BlokShifman:93,BlokShifman2:93,BlokShifman:92}. 
For semi-leptonic decays NLO-QCD corrections are know as 
well~\cite{Gambino:14,Mannel:2014xza,Mannel:2015jka}.

The contribution from the second power correction due to the dimension six Darwin and spin-orbit operators is known at LO-QCD~\cite{GremKapustin:97}
and NLO-QCD~\cite{Mannel:2019qel} only for the semi-leptonic case. 
However, the $\rho_D$ contribution for 
inclusive non-leptonic decays, is still missing. That is precisely the task we address in this this work.
In fact, previous studies focus on the four-quark operators appearing at this order
which induce lifetime differences at tree level and which are parametrically enhanced 
by a phase space factor $16\pi^2$. However, our explicit calculation shows that the coefficient in front of 
$\rho_D$ turns out to be enhanced and thus needs to be taken into account. 
The contribution from dimension six four-quark operators is known at NLO-QCD~\cite{BBGLN:02,Mescia:2002,Lenz:2013}.

\section{Outline of the calculation}
\label{Sec:Outline}
In this section we give a brief outline of the calculation which in fact contains 
a few subtleties. A detailed description will be deferred to a more technical 
publication. We start from the effective Lagrangian
for flavor changing transitions due to charged hadronic 
currents~\cite{BBL:96}
\begin{equation} \label{Leff}
 \mathcal{L}_{{\footnotesize{\mbox{eff}}}} = - \frac{4G_F}{\sqrt{2}}V_{CKM}' V_{CKM}^* 
(C_1 \mathcal{O}_1 + C_2 \mathcal{O}_2) 
 + \mbox{h.c}\,,
\end{equation}
where $G_F$ is the Fermi constant, ${\cal O}_{1,2}$ are four-quark operators,
and $V_{CKM}, V_{CKM}'$ are the corresponding Cabibbo-Kobayashi-Maskawa 
matrix elements describing weak mixing of quark generations.
We consider only the tree-level four-quark operators of current-current type
since the Wilson coefficients of these operators are the largest.  
The numerical values of the Wilson coefficients $C_{1,2}(\mu)$ at the 
scale $\mu=m_b$, where $m_b$ is the value of the $b$-quark mass, 
are known in the SM with 
high precision mainly thanks to using the high order 
renormalization group improved QCD perturbation theory at the scales 
between $m_b$ and $M_W$.

We are interested in weak decays of beauty hadrons mediated by the CKM leading 
transitions with the flavour structure  $b\to c \bar u d$ and $b\to c\bar c s$.
The latter decay is additionally slightly suppressed by the phase space 
available for the decay products due to the mass of the $c$-quarks.
The canonical choice of the operator basis for the decays $b\to c\bar q_1 q_2$ 
reads~\cite{BBL:96}
\begin{equation}
\label{eq:canbasis}
 \mathcal{O}_1 = (\bar c^i_L \gamma_\mu b^i_L)(\bar q_{2\,L}^j \gamma^\mu q_{1\,L}^j)\,,\quad
 \mathcal{O}_2 = (\bar c^i_L \gamma_\mu b^j_L)(\bar q_{2\,L}^j \gamma^\mu q_{1\,L}^i)\,,
\end{equation}
where $q_L$ denotes the left-handed quark.
It is the basis~(\ref{eq:canbasis}) that is used for the computation
of the Wilson coefficients. 

However, for the purposes of the present computation we use a different operator basis (cf.~\cite{Beneke:2002rj}), which is obtained 
after applying a four-dimensional Fierz transformation to $\mathcal{O}_2$. The operators of the new basis 
are diagonal in the color space and have the form
\begin{equation}
 \mathcal{O}_1 = (\bar c^i \Gamma_\mu b^i)(\bar q_2^j \Gamma^\mu q_1^j)\,,\quad
 \mathcal{O}_2 = (\bar q_2^i \Gamma_\mu b^i)(\bar c^j \Gamma^\mu q_1^j) \,,
\end{equation}
with $\Gamma_\mu=\gamma_\mu (1-\gamma_5)/2$. 
We consider two Cabibbo favoured 
decay channels, $(q_1,q_2) = (u,d)$ and $(q_1,q_2) = (c,s)$ . 

The main technical tool for our computation is dimensional regularization 
($D=4-2\epsilon$)~\cite{tHooft:1972tcz}.
The Dirac algebra of $\gamma$-matrices is usually defined in $D=4$ and needs to 
be properly extended to $D$-dimensional space 
time~\cite{Altarelli:1980fi,Buras:1989xd,Buras:1990fn,Chanowitz:1979zu}.
In particular, using Fierz transformations can lead to a non-trivial $\epsilon$ 
dependence~\cite{Dugan:1990df,Herrlich:1994kh,Pivovarov:1988gt,Pivovarov:1991nk}. 
With this in mind, an arbitrary change of the operator basis valid in four-dimensional 
space is not allowed if perturbative corrections of higher order are to be included:
the change will require the corresponding change of 
the set of evanescent operators associated with a given 
basis. For our computation however the required accuracy 
is such that one can use the new basis without changing the coefficients $C_{1,2}$. 
A review of the relevant techniques can be found in, 
e.g.~\cite{Chetyrkin:1997gb,Grozin:2017uto}.

The $B$ meson decay rate for the inclusive non-leptonic decays 
can be computed from the discontinuity of the forward scattering
matrix element which is computed in the HQE.
The property of unitarity of the $S$-matrix and the optical 
theorem lead to an expression for the decay width in the form 
\begin{eqnarray}
 \Gamma(b\rightarrow c \bar q_1 q_2) &=& 
 \frac{1}{2M_B} \langle B(p_B)\lvert \mbox{Im}\, i \int d^4 x\, 
T\{\mathcal{L}_{\mbox{\scriptsize eff}}(x),\mathcal{L}_{{\scriptsize\rm eff}}(0)\} \lvert B(p_B)\rangle
\nonumber
\\
 &=& \frac{1}{2M_B}\langle B(p_B)\lvert \mbox{Im}\,\mathcal{T} \lvert B(p_B)\rangle\, .
\end{eqnarray}
The HQE of the transition operator up to $1/m_b^3$ is given by
\begin{equation}
\label{hqeTOPm3}
 \mbox{Im}\, \mathcal{T} = \Gamma_{\bar q_1 q_2}^0
 \bigg( C_0 \mathcal{O}_0 
 + C_v \frac{\mathcal{O}_v}{m_b} 
 + C_\pi \frac{\mathcal{O}_\pi}{2m_b^2} 
 + C_G \frac{\mathcal{O}_G}{2m_b^2} 
 + C_D \frac{\mathcal{O}_D}{4m_b^3}
 + C_{LS} \frac{\mathcal{O}_{LS}}{4m_b^3}
 + \sum_{i,q} C_{4F_i}^{(q)} \frac{\mathcal{O}_{4F_i}^{(q)} }{4m_b^3}
 \bigg)\,.
\end{equation} 
Here $\Gamma_{\bar q_1 q_2}^0 = G_F^2 m_b^5 |V_{cb}|^2 |V_{q_1 q_2}|^2/(192\pi^3)$, $V_{cb}$ and $V_{q_1 q_2}$ are the corresponding CKM matrix elements, 
$q$ stands for a massless quark and
\begin{eqnarray}
 \mathcal{O}_0 &=& \bar h_v h_v\,,
\\
 \mathcal{O}_v &=& \bar h_v (v\cdot \pi) h_v\,,
\\
 \mathcal{O}_\pi &=& \bar h_v \pi_\perp^2 h_v\,,
\\
 \mathcal{O}_G &=& \frac{1}{2}\bar h_v[\slashed \pi_{\perp}, \slashed \pi_{\perp}] h_v 
 = \frac{1}{2}\bar h_v [\gamma^\mu, \gamma^\nu] \pi_{\perp\,\mu}\pi_{\perp\,\nu}  h_v\,,
\\
 \mathcal{O}_D &=& \bar h_v[\pi_{\perp\,\mu},[\pi_{\perp}^\mu , v\cdot \pi]] h_v\,,
\\
 \mathcal{O}_{LS} &=& \frac{1}{2}\bar h_v[\gamma^\mu,\gamma^\nu]\{ \pi_{\perp\,\mu},[\pi_{\perp\,\nu}, v\cdot \pi] \} h_v\,,
\end{eqnarray}
are HQET local operators with $\pi_\mu= iD_\mu = i\partial_\mu +g_s A_\mu^a T^a$ and $\pi^\mu = v^\mu (v\pi) + \pi_\perp^\mu$. The four-quark operators 
$\mathcal{O}_{4F_i}^{(q)}$ will be defined in Sec. \ref{Sec:match4F}.

Finally, the QCD spinor $b$ of the bottom quark is replaced by the HQET fermion field $h_v$. They are 
related as follows
\begin{equation}
 b = e^{-im_b v\cdot x}\bigg[
 1 + \frac{\slashed \pi_\perp}{2m_b} 
 - \frac{(v\cdot \pi)\slashed \pi_\perp}{4m_b^2} 
 + \frac{\slashed\pi_\perp \slashed\pi_\perp}{8m_b^2} 
 + \frac{(v\cdot \pi)^2 \slashed \pi_\perp}{8m_b^3}
 + \frac{\slashed \pi_\perp \slashed\pi_\perp \slashed\pi_\perp}{16m_b^3}
 + \mathcal{O}(1/m_b^4)
 \bigg]h_v\,.
\end{equation}

\subsection{Matching of four-fermion operators: computation of $C_{4F_i}^{(q)}$}
\label{Sec:match4F}
\begin{figure}[!htb]  
	\centering
	\includegraphics[width=1.0\textwidth]{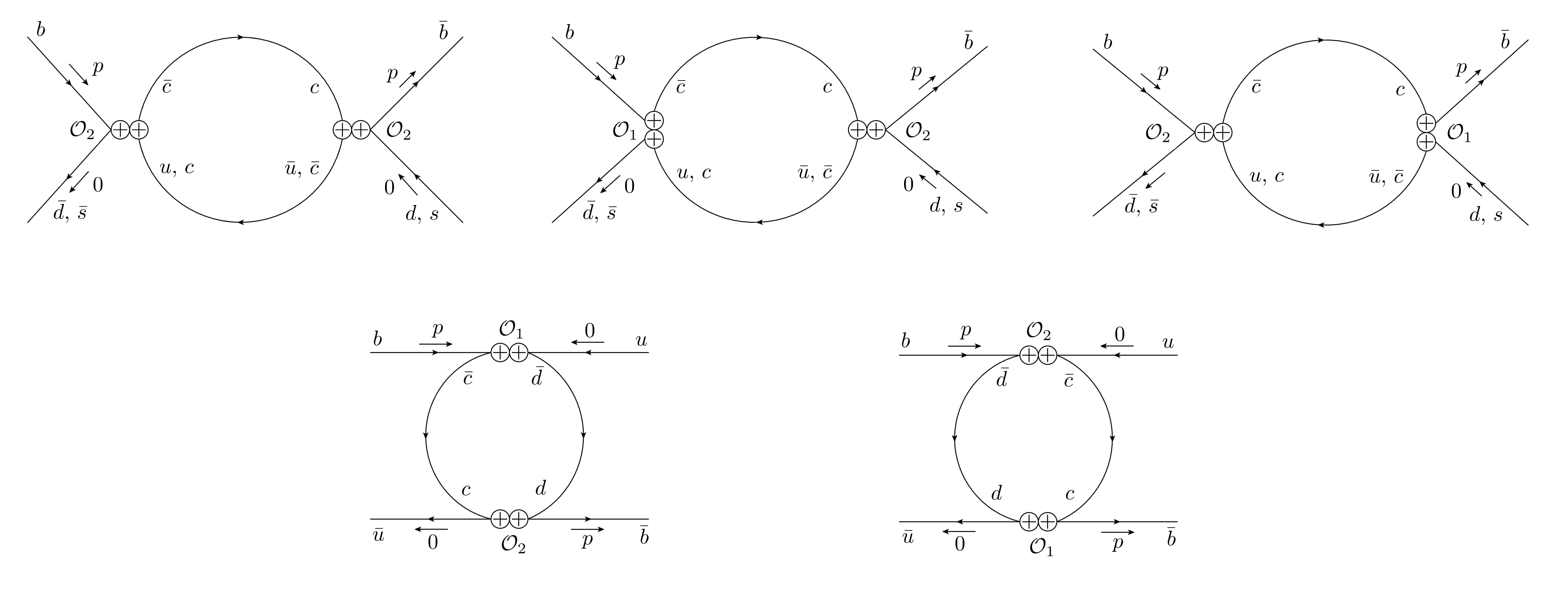}
\caption{One loop diagrams contributing to the matching coefficients of four-quark operators in the HQE of the forward scattering
matrix element of the $B$ meson.}
\label{Matching4Fop}   
\end{figure}

In this subsection we give some sketch of the matching calculation of four-quark operators relevant for the 
renormalization of the coefficient $C_D$. We chose the version of the HQE where the bottom 
and the charm quarks are integrated out at the same scale $m_c \le \mu \le m_b$ such 
that only the $u,d,s$ quarks remain as soft (massless) dynamical quarks. 
As a consequence, the matching coefficients will depend on the mass ratio 
$r=m_c^2/m_b^2 \sim {\cal O}(1)$. We only compute those pieces which are 
relevant for the renormalization of $C_D$. Such contributions are diagramatically represented in Fig.~[\ref{Matching4Fop}].

\subsubsection{The channel $b\rightarrow c\bar u d$}
\label{SubSec:match4qbcud}
The relevant operators in the HQE are
\begin{eqnarray}
 \mathcal{O}_{4F_1}^{(d)} &=& (\bar h_v \Gamma_\mu d) (\bar d \Gamma^\mu h_v)\,, \\ 
 \mathcal{O}_{4F_2}^{(d)} &=& (\bar h_v P_L d) (\bar d P_R h_v)\,, \\
\mathcal{O}_{4F_1}^{(u)} &=& (\bar h_v \Gamma^\sigma \gamma^\mu \Gamma^\rho u) (\bar u \Gamma_\sigma \gamma_\mu \Gamma_\rho h_v)\,, \\ 
 \mathcal{O}_{4F_2}^{(u)} &=& (\bar h_v \Gamma^\sigma \slashed v \Gamma^\rho u) (\bar u \Gamma_\sigma \slashed v \Gamma_\rho h_v)\,,
\end{eqnarray}
with the matching coefficients in $D=4-2\epsilon$ dimensions  
\begin{eqnarray}
   C_{4F_1}^{(d)} \!\! &=& \!\!\!\!
  -(3C_2^2 + 2C_1 C_2 (1-\epsilon) )\frac{3\cdot 2^{6+4\epsilon} \pi^{5/2+\epsilon} m_b^{-2\epsilon} (1-r)^{2-2\epsilon}(2+r-2\epsilon)}{\Gamma(5/2-\epsilon)} \\
  \!\! &=& \!\!\!\!  -(3C_2^2 + 2C_1 C_2 ) 256 \pi^2 (1-r)^2 (2+r) \quad \mbox{for $\epsilon \to 0$}\,, 
  \nonumber \\
  C_{4F_2}^{(d)} \!\!&=& \!\!\!\!
 -(3C_2^2 + 2C_1 C_2(1-\epsilon) )
 \frac{3\cdot 2^{7+4\epsilon} \pi^{5/2+\epsilon} m_b^{-2\epsilon} (1-r)^{2-2\epsilon} (-1+r(-2+\epsilon)+\epsilon)}{ \Gamma(5/2-\epsilon)} \\
 \!\! &=& \!\!\!\! (3C_2^2 + 2C_1 C_2 ) 512 \pi^2 (1-r)^2 (1+2r) \quad \mbox{for $\epsilon \to 0$}\,,     \nonumber \\
  C_{4F_1}^{(u)} \!\!&=& \!\!\!\! C_1 C_2 \frac{3\cdot 2^{5+4\epsilon}  m_b^{-2\epsilon} \pi^{5/2+\epsilon}(1-r)^{3-2\epsilon}}{\Gamma(5/2-\epsilon)} \\
   \!\! &=& \!\!\!\! C_1 C_2  128 \pi^2 (1-r)^3 \quad \mbox{for $\epsilon \to 0$}\,,  \nonumber \\
 C_{4F_2}^{(u)} \!\!&=& \!\!\!\! - C_1 C_2 \frac{3\cdot 2^{6+4\epsilon}\pi^{5/2+\epsilon}  m_b^{-2\epsilon} (1-r)^{2-2\epsilon} (-1+r(-2+\epsilon)+\epsilon)} 
 {\Gamma(5/2-\epsilon)} \\
  \!\! &=& \!\!\!\! C_1 C_2 256 \pi^2 (1-r)^2 (1+2r) \quad  \mbox{for $\epsilon \to 0$}\,.   \nonumber 
\end{eqnarray}
%
These expressions coincide with the results 
of ref.~\cite{Beneke:2002rj}.

\begin{figure}[!htb]  
	\centering
	\includegraphics[width=0.3\textwidth]{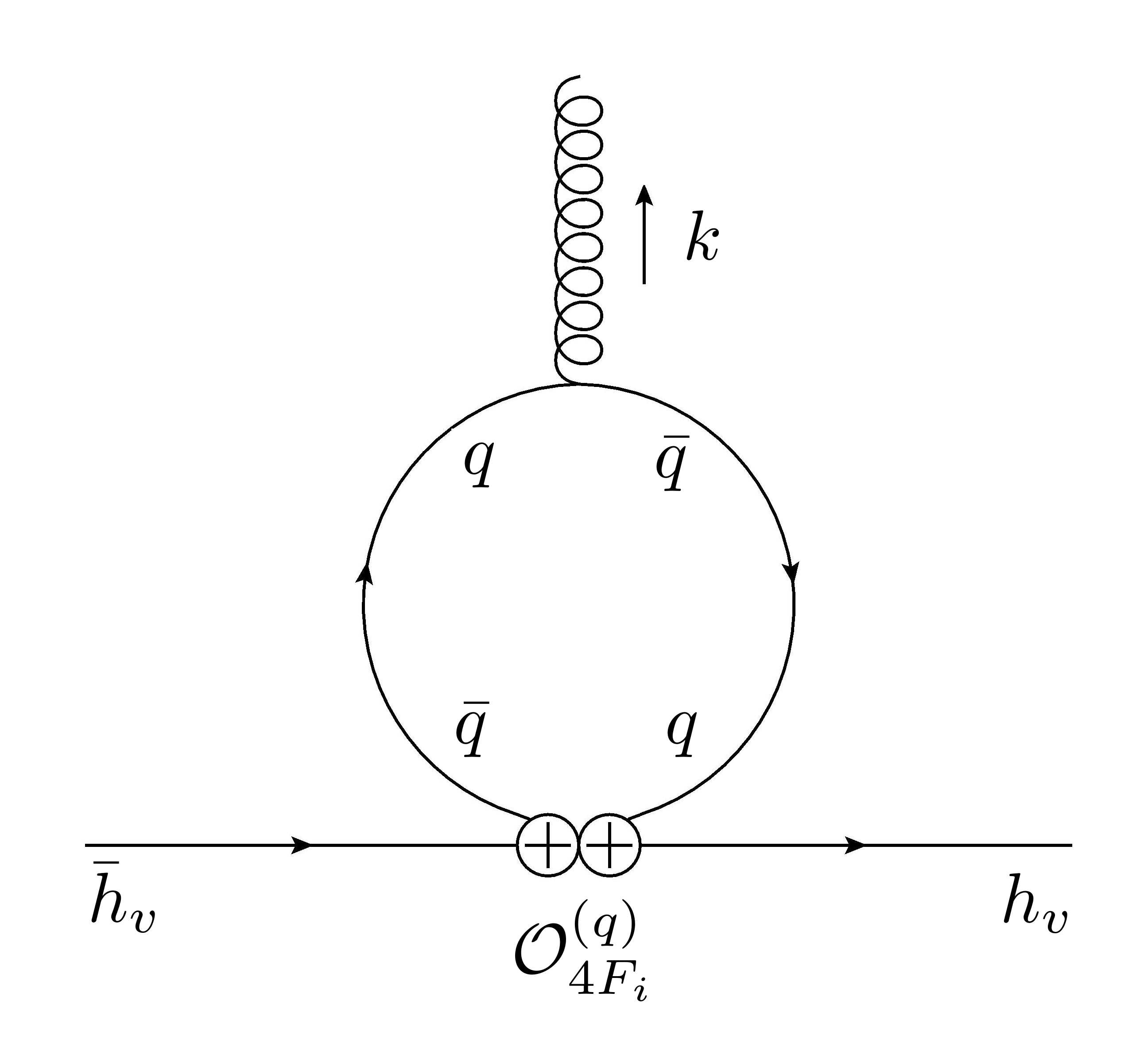}
\caption{One loop diagrams contributing the renormalization of $C_{D}$.}
\label{ren}   
\end{figure}

The one-loop matrix elements of these four-fermion operators also contribute to 
the coefficient $C_D$ (see Fig.~\ref{ren}). We find that

\begin{equation}
 \mbox{Im}\, \mathcal{T} =  \ldots +
 \Gamma_{ud}^0  \frac{1}{4m_b^3} \bigg(- C_{4F_1}^{(d)} + \frac{1}{2}C_{4F_2}^{(d)} -16C_{4F_1}^{(u)}-4C_{4F_2}^{(u)}\bigg)
 \frac{1}{48\pi^2\epsilon}(-k^2)^{-\epsilon}  \mathcal{O}_D\,,
\end{equation}
and we can determine the counterterm of 
$C_D$ in the $\overline{\mbox{MS}}$ renormalization scheme. We obtain
\begin{equation}
 \delta C_D^{{\scriptsize\overline{\mbox{MS}}}}(\mu) = 
 \bigg( C_{4F_1}^{(d)} - \frac{1}{2}C_{4F_2}^{(d)} + 16C_{4F_1}^{(u)} + 4C_{4F_2}^{(u)}\bigg)
 \frac{1}{48\pi^2\epsilon}\mu^{-2\epsilon}\bigg(\frac{e^{\gamma_E}}{4\pi}\bigg)^{-\epsilon}\,,
\end{equation}
where $C_D^B = C_D^{{\scriptsize\overline{\mbox{MS}}}}(\mu) + \delta C_D^{{\scriptsize\overline{\mbox{MS}}}}(\mu)$ and
$\bar\mu^{-2\epsilon} = \mu^{-2\epsilon}( e^{\gamma_E}/4\pi)^{-\epsilon}$ 
is the $\overline{\mbox{MS}}$ renormalization scale.

\subsubsection{The channel $b\rightarrow c\bar c s$}

The relevant four-quark operators of the HQE are
\begin{eqnarray}
 \mathcal{O}_{4F_1}^{(s)} &=& (\bar h_v \Gamma_\mu s) (\bar s \Gamma^\mu h_v)\,,
\\
 \mathcal{O}_{4F_2}^{(s)} &=& (\bar h_v P_L s) (\bar s P_R h_v)\,,
\end{eqnarray}
with matching coefficients (see Fig.~\ref{Matching4Fop})
\begin{eqnarray}
 C_{4F_1}^{(s)}  &=& 
  - (3C_2^2 + 2C_1 C_2(1-\epsilon)) \frac{3\cdot 2^{7+4\epsilon}m_b^{-2\epsilon}\pi^{5/2+\epsilon}z^{1-2\epsilon}(1-\epsilon + r(-1+2\epsilon))}{\Gamma(5/2-\epsilon)}
\\
  \!\! &=& \!\!\!\! - (3C_2^2 + 2C_1 C_2) 512 \pi^2 z (1 - r) \quad  \mbox{for $\epsilon \to 0$}\,,   \nonumber 
  \\
  C_{4F_2}^{(s)} &=& 
 - (3C_2^2 + 2C_1 C_2(1-\epsilon) ) \frac{3\cdot 2^{7+4\epsilon} m_b^{-2\epsilon}\pi^{5/2+\epsilon}z^{1-2\epsilon}(-1-2r+\epsilon)}{\Gamma(5/2-\epsilon)}
 \\
  \!\! &=& \!\!\!\!  (3C_2^2 + 2C_1 C_2 ) 512 \pi^2 z (1+2r) \quad  \mbox{for $\epsilon \to 0$}\,.   \nonumber 
\end{eqnarray}
Again the one-loop matrix elements of the four-fermion operators also contribute to $C_D$ (see Fig. [\ref{ren}]). 
We find that
\begin{equation}
 \mbox{Im}\, \mathcal{T} = \ldots + 
 \Gamma_{cs}^0\frac{1}{4m_b^3} \bigg(
  -C_{4F_1}^{(s)} 
 + \frac{1}{2} C_{4F_2}^{(s)}  
 \bigg)\frac{1}{48\pi^2\epsilon}(-k^2)^{-\epsilon} \mathcal{O}_D\,,
\end{equation}
and we can determine the counterterm of $C_D$ in the $\overline{\mbox{MS}}$ 
renormalization scheme, for which we obtain
\begin{equation}
 \delta C_D^{{\scriptsize\overline{\mbox{MS}}}}(\mu) = 
 \bigg(
  C_{4F_1}^{(s)} 
 - \frac{1}{2} C_{4F_2}^{(s)}  
 \bigg)\frac{1}{48\pi^2\epsilon}\mu^{-2\epsilon}\bigg(\frac{e^{\gamma_E}}{4\pi}\bigg)^{-\epsilon}\,,
\end{equation}
where $C_D^B = C_D^{{\scriptsize\overline{\mbox{MS}}}}(\mu) + \delta C_D^{{\scriptsize\overline{\mbox{MS}}}}(\mu)$ and
$\bar\mu^{-2\epsilon} = \mu^{-2\epsilon}( e^{\gamma_E}/4\pi)^{-\epsilon}$ 
is the $\overline{\mbox{MS}}$ renormalization scale.

\subsection{Matching of two-fermion operators: computation of $C_i$}
\label{Sec:match2F}

\begin{figure}[!htb]  
	\centering
	\includegraphics[width=1.0\textwidth]{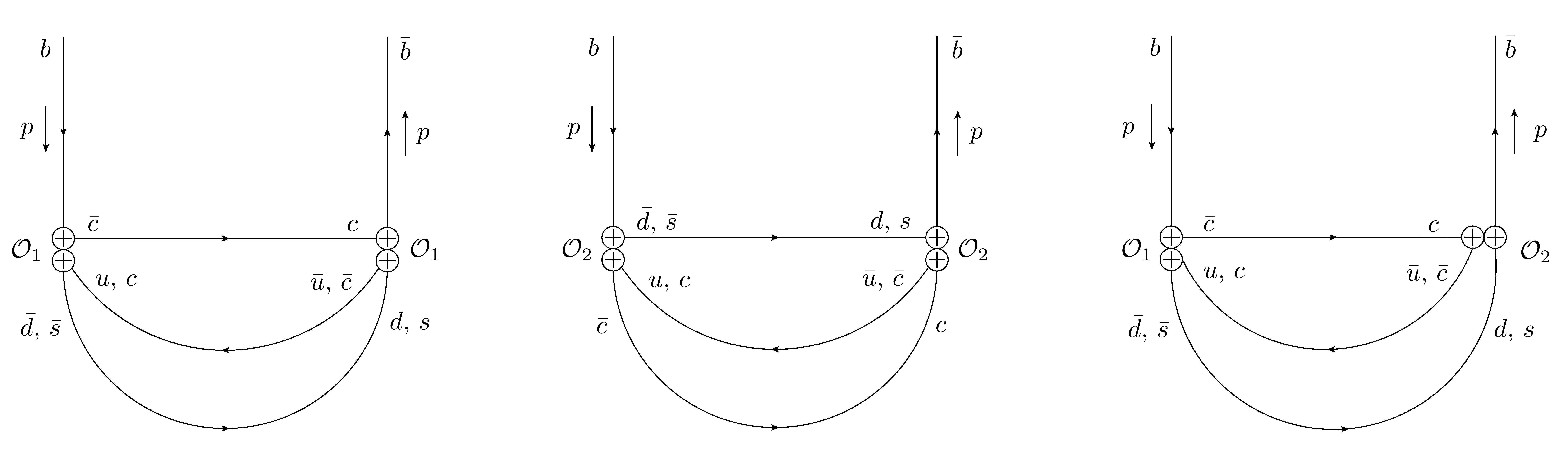}
\caption{Two loop diagrams contributing to the matching coefficients of two-quark operators in the HQE of the forward scattering
matrix element of the $B$ meson.}
\label{forwardsme}   
\end{figure}

In this section we describe the matching computation of two-quark operators. The different contributions are diagramatically represented in 
Fig.~[\ref{forwardsme}]. In order to optimize the computation we find expressions for the quark propagator in an external gluon field $A$.

In the semi-leptonic case the tree level expression for the HQE can be obtained from 
expanding the external field propagator for the charm or the up quark in powers of 
the covariant derivative
\begin{equation}
S_q = \frac{i}{\slashed{Q} + \slashed{\pi} - m } =  \frac{i}{\slashed{Q}  - m } 
\sum_{\nu = 0}^\infty 
\left[  i\slashed{\pi} \frac{i}{\slashed{Q}  - m }  \right]^\nu \,,
\end{equation}   
which automatically generates the proper ordering of the covariant derivatives. 
However, in the non-leptonic case the leptonic lines are replaced by quark lines 
and hence we pick up additional diagrams where gluons are emitted from these quark 
lines and the simple trick from the semi-leptonic calculation cannot be used here. 

Still such contributions can most easily be taken into 
account by using the expression of the quark propagator in the external 
gluon field in the Fock-Schwinger gauge 
$x^\mu A_\mu(x) = 0$ (see, e.g.~\cite{NSVZ:84}). This is 
especially convenient because the expansion of the propagator has then an explicitly gauge invariant form. Another important 
property of the Fock-Schwinger gauge is that it breaks explicitly the translation invariance of the quark propagator, namely 
$S_F(x,0)\neq S_F(0,x)$. We obtain
\begin{eqnarray}
S_F(x,0) = \int\frac{d^4 p}{(2\pi)^4}e^{-ipx}S_F(p)\,,\quad S_F(0,x) = \int\frac{d^4 p}{(2\pi)^4}e^{ipx}\tilde S_F(p)\,,
\end{eqnarray}
with explicit expressions in momentum space given by
\begin{eqnarray}
 S_F(p)  &=& S_F^{(0)}(p) + \frac{1}{2}[\pi_\rho, \pi_\sigma]  S_F^{(0)}(p) i\gamma^\rho S_F^{(0)}(p) i\gamma^\sigma S_F^{(0)}(p)
\\
&&
 + \frac{1}{3} ( [\pi_\sigma, [\pi_\rho, \pi_\lambda]] + [\pi_\rho, [\pi_\sigma, \pi_\lambda]]) 
  S_F^{(0)}(p)  i\gamma^\lambda S_F^{(0)}(p) i\gamma^\sigma S_F^{(0)}(p) i\gamma^\rho S_F^{(0)}(p)\,,
  \nonumber
\end{eqnarray}

\begin{eqnarray}
 \tilde S_F(p)  &=& S_F^{(0)}(p) + \frac{1}{2}[\pi_\rho, \pi_\sigma] S_F^{(0)}(p) i\gamma^\rho S_F^{(0)}(p) i\gamma^\sigma  S_F^{(0)}(p)
\\
&&
 +  \frac{1}{3}( [\pi_\lambda, [\pi_\sigma, \pi_\rho]] + [\pi_\sigma, [\pi_\lambda, \pi_\rho]])
 S_F^{(0)}(p) i\gamma^\lambda S_F^{(0)}(p) i\gamma^\sigma S_F^{(0)}(p) i\gamma^\rho S_F^{(0)}(p)\,,
\nonumber
\end{eqnarray} 
where $S_F^{(0)}(p)$ is the free quark propagator

\begin{equation}
 S_F^{(0)}(p) = \frac{i(\slashed p + m)}{p^2 - m^2}\,.
\end{equation}
The expressions $S_F(p)$ and $\tilde S_F(p)$ are used for the propagator of the $\bar q_1$-quark and $q_2$-quark respectively 
to compute the diagrams that do not appear in the semi-leptonic case.  

Let's us discuss the peculiarities of each contribution. The computation of the 
$\mathcal{O}_1\otimes\mathcal{O}_1$ contribution goes exactly as in the semi-leptonic case. The color structure of the operator
$\mathcal{O}_1$ only allows 
for the radiation of a single gluon from the $c$-quark in the 
${\bar b} S_c b$ line. So we only need to expand the $c$-quark propagator.
The computation is then identical to the case of semi-taonic decays
and the corresponding results can be taken from ref.~\cite{Mannel:2017jfk}.

The computation of $\mathcal{O}_2\otimes\mathcal{O}_2$ proceeds as the semi-leptonic case as well after replacing $c\rightarrow q_2$. 
The color structure only allows radiation of a single gluon 
from the $q_2$-quark in the 
$b S_{q_2} b$ line ($q_2 = d,s$), so we only need to expand this $q_2$-quark propagator.
In this case one faces the 
IR divergences due to the gluon emission or the expansion of the massless 
quark propagator. Within the HQE (and OPE in general) the appearance of 
such infrared divergence signals the mixing between the local operators 
that constitute the basis of the expansion. The corresponding local operator 
develop UV divergences and should be properly renormalized.
The well known advantage of using dimensional regularization is that both IR and UV divergences are dealt with simultaneously and a uniform manner.
This treatment allows us to retain some vital symmetries of the theory and has technical superiority  of simplicity.
In fact, it is just this phenomenon of mixing that is the most essential 
and interesting part of the whole calculation. 

The computation of $\mathcal{O}_1\otimes\mathcal{O}_2$, which is found to be 
the same that for $\mathcal{O}_2\otimes\mathcal{O}_1$, differs    
from the one in the semi-leptonic case. Here the gluon emission or the expansion 
of the quark propagators from all lines have to be taken into account. Overall, 
the computation of the coefficient of the $\rho_{LS}$ operator in HQE is infrared 
safe even for massless quarks, does not require considering mixing with four 
quark operators,  and can be performed in $D=4$.

\section{Results for the Wilson coefficients at order $1/m_Q^3$}
\label{sect:results}
Before we give our results for the terms of order $1/m_b^3$, 
we need to discuss the effects induced by operator mixing. 
The HQE as any OPE of effective theory
gives an example of the general phenomenon of the
separation of physics at greatly different scales.
Indeed, the hadronic width in the representation
\begin{eqnarray}
\Gamma(b\rightarrow c \bar q_1 q_2) &=&
\frac{1}{2M_B}\mbox{Im}\, \langle B(p_B)\lvert i \int d^4 x\,
T\{\mathcal{L}_{\mbox{\scriptsize eff}}(x),\mathcal{L}_{{\scriptsize\rm
		eff}}(0)\} \lvert B(p_B)\rangle
\nonumber
\\
&=& \frac{1}{2M_B}\langle B(p_B)\lvert \mbox{Im}\,\mathcal{T} \lvert
B(p_B)\rangle\, ,
\end{eqnarray}
depends on the heavy quark mass $m_b$ and the infrared scale of QCD
$\Lambda_{\rm QCD}$ with $m_b\gg \Lambda_{\rm QCD}$.
The HQE in expression~(\ref{hqeTOPm3}) is organized in such a
way that the coefficients are insensitive to
$\Lambda_{\rm QCD}$ while the matrix elements of the local operators
are independent of the large scale $m_b$.
An explicit naive computation, however, produces at some intermediate
stage
both IR singularities in the coefficient functions and UV singularities
in the matrix elements. The combinatorics of HQE is such that
all singularities cancel. Technically the most efficient way to perform
computations is to use dimensional regularization for both IR and UV
divergences. In such a setup one has to take into account the mixing of
local operators at UV renormalization.
Thus, an infrared divergence of coefficient functions
signals the UV mixing between the local operators
that constitute the basis of the expansion.
The corresponding local operator
develop then UV divergences and should be properly renormalized.

In our computation a naive way of getting the coefficient of the
$\rho_D$ operator leads to an IR singularity in
the contribution of $\mathcal{O}_2\otimes\mathcal{O}_2$
and $\mathcal{O}_1\otimes\mathcal{O}_2$ correlators
as one gets the radiation of a soft gluon
from the light quark line.

This singularity is canceled by the UV
renormalization of the four-quark operator
that has the general form (quite symbolically)
\begin{equation}
\label{eq:gen4rhomixi}
\langle{\bar b} \gamma s {\bar s} \gamma b\rangle^{\rm ren}
= \langle{\bar b} \Gamma s {\bar s} \Gamma b\rangle^{\rm B}
+\gamma(\Gamma)\frac{1}{\epsilon}\rho_D\,,
\end{equation}
where the matrix $\Gamma$ gives the corresponding Dirac structure of
the four-quark operator and the quantity $\gamma(\Gamma)$ is the mixing
anomalous dimension depending on the Dirac structure $ \Gamma$.
The UV pole in $\epsilon$ coming from the operator mixing in
eq.~(\ref{eq:gen4rhomixi}) cancel the IR divergence in the
coefficient function for the operator $\rho_D$.

The HQE of the imaginary part of the transition 
operator is given by eq.~(\ref{hqeTOPm3}). 
However, it is convenient to rewrite it in terms of 
the local operator ${\bar b}\slashed v b$ defined in full QCD. It can be employed 
to remove $\mathcal{O}_0$ in the HQE by using the expansion
\begin{equation}
\bar b \slashed v b = \mathcal{O}_0 - \tilde C_\pi \frac{\mathcal{O}_\pi}{2m_b^2} 
+ \tilde C_G \frac{\mathcal{O}_G}{2m_b^2} 
 + \tilde C_D \frac{\mathcal{O}_D}{4m_b^3}
 + \tilde C_{LS} \frac{\mathcal{O}_{LS} }{4m_b^3}
 + \mathcal{O}(\Lambda_{QCD}^4/m_b^4)\,.
\end{equation}
Inserting this expression 
we get (omitting here and in what it follows the four-quark contributions)
\begin{eqnarray}
  \mbox{Im} \mathcal{T} &=& \Gamma_{\bar q_1 q_2}^0 
 \bigg[ C_0 \bigg(\bar b \slashed v b 
 + \tilde C_\pi \frac{\mathcal{O}_\pi}{2m_b^2} 
 - \tilde C_G \frac{\mathcal{O}_G}{2m_b^2} 
 - \tilde C_D \frac{\mathcal{O}_D}{4m_b^3}
 - \tilde C_{LS} \frac{\mathcal{O}_{LS} }{4m_b^3}\bigg) 
 \nonumber
 \\
 &&
 + C_v \frac{\mathcal{O}_v}{m_b} 
 + C_\pi \frac{\mathcal{O}_\pi}{2m_b^2} 
 + C_G \frac{\mathcal{O}_G}{2m_b^2} 
 + C_D \frac{\mathcal{O}_D}{4m_b^3}
 + C_{LS} \frac{\mathcal{O}_{LS}}{4m_b^3}
 \bigg]\,,
\end{eqnarray}
which has the advantage that the forward matrix element of the leading term 
is normalized to all orders. 

For the operator  $\mathcal{O}_v$ we use the equation of motion
\begin{equation}
 \mathcal{O}_v =
 - \frac{1}{2m_b} (\mathcal{O}_\pi+ C_{{\scriptsize\mbox{mag}}}(\mu)\mathcal{O}_G)
 -  \frac{1}{8m_b^2} (c_D(\mu)\mathcal{O}_D + c_S(\mu)\mathcal{O}_{LS})
\end{equation}
to remove it from the expression for $ \mbox{Im}\,\mathcal{T}$

\begin{eqnarray}
 \mbox{Im}\,\mathcal{T} &=& \Gamma_{\bar q_1 q_2}^0
 \bigg[ C_0 \bigg( \bar b \slashed v b 
+ \frac{C_\pi  + C_0 \tilde C_\pi -  C_v}{C_0}\frac{\mathcal{O}_\pi}{2m_b^2}\bigg)
+ \bigg( C_G - C_0 \tilde C_G -  C_v C_{{\scriptsize\mbox{mag}}}(\mu) \bigg) \frac{\mathcal{O}_G}{2m_b^2}
\nonumber
\\
&&
+ \bigg(C_D- C_0 \tilde C_D-\frac{1}{2} C_v c_D(\mu) \bigg) \frac{\mathcal{O}_D}{4m_b^3}
 + \bigg( C_{LS}- C_0 \tilde C_{LS}-  \frac{1}{2} C_v c_S(\mu) \bigg) \frac{\mathcal{O}_{LS}}{4m_b^3}
 \bigg]\,.
 \nonumber
\\
&&
\end{eqnarray}
This is the desired expression for the transition operator, from which we compute 
the total decay rate
\begin{equation}
 \Gamma(b\rightarrow c \bar q_1 q_2) = 
 \frac{1}{2M_B}\langle B(p_B)\lvert \mbox{Im}\,\mathcal{T} \lvert B(p_B)\rangle
\end{equation}
in terms of the HQE parameters
\begin{eqnarray}
 \langle B(p_B)\lvert \bar b \slashed v b \lvert B(p_B)\rangle &=& 2M_B\,,  \\
 - \langle B(p_B)\lvert \mathcal{O}_\pi \lvert B(p_B)\rangle &=& 2M_B \mu_\pi^2\,, \\ 
 C_{{\scriptsize\mbox{mag}}}(\mu)\langle B(p_B)\lvert \mathcal{O}_G \lvert B(p_B)\rangle
 &=& 2M_B \mu_G^2\,, \\
  - c_D(\mu)\langle B(p_B)\lvert \mathcal{O}_D \lvert B(p_B)\rangle&=& 4M_B \rho_D^3\,, \\
 -  c_S(\mu)\langle B(p_B)\lvert \mathcal{O}_{LS} \lvert B(p_B)\rangle&=& 4 M_B \rho_{LS}^3\,.
\end{eqnarray}
%
We obtain 

\begin{eqnarray}
 \Gamma(b\rightarrow c \bar q_1 q_2) &=& \Gamma_{\bar q_1 q_2}^0 
 \bigg[ C_0 \bigg( 1 
- \frac{C_\pi  + C_0 \tilde C_\pi -  C_v}{C_0}\frac{\mu_\pi^2}{2m_b^2}\bigg)
+ \bigg(\frac{C_G - C_0 \tilde C_G}{C_{{\scriptsize\mbox{mag}}}(\mu)}-  C_v \bigg)\frac{\mu_G^2}{2m_b^2}
\nonumber
\\
&&
 - \bigg(\frac{C_D- C_0 \tilde C_D}{c_D(\mu)}-\frac{1}{2} C_v\bigg) \frac{\rho_D^3}{2m_b^3}
 - \bigg(\frac{C_{LS}- C_0 \tilde C_{LS} }{c_S(\mu)} - \frac{1}{2} C_v \bigg) \frac{\rho_{LS}^3}{2m_b^3}
 \bigg]\,.
\end{eqnarray}
At leading order we have  $c_S=c_D=C_{{\scriptsize\mbox{mag}}}=1$. 
It is convenient to define new coefficients corresponding to every matrix element
\begin{equation} \label{Res1} 
 \Gamma(b\rightarrow c \bar q_1 q_2) = \Gamma_{\bar q_1 q_2}^0 
 \bigg[   C_0 
- C_{\mu_\pi}\frac{\mu_\pi^2}{2m_b^2}
+ C_{\mu_G}\frac{\mu_G^2}{2m_b^2}
- C_{\rho_D} \frac{\rho_D^3}{2m_b^3}
 - C_{\rho_{LS}} \frac{\rho_{LS}^3}{2m_b^3}
 \bigg]\,.
\end{equation}
We find for the coefficients in the case of $\Gamma(b\rightarrow c\bar u d)$


\begin{eqnarray}
 &C_0& = C_{\mu_\pi} = (3 C_1^2 + 2 C_1 C_2 + 3 C_2^2) (1 - 8 r + 8 r^3 - r^4 - 12 r^2 \ln(r) )\,,
 \\
 &&
 \nonumber
 \\
 &C_v& = (3 C_1^2 + 2 C_1 C_2 + 3 C_2^2) (5 - 24 r + 24 r^2 - 8 r^3 + 3 r^4 - 12 r^2 \ln(r) )\,,
  \\
   &&
   \nonumber
 \\
 &C_{\mu_G}& = C_{\rho_{LS}} = 3(C_1^2 + C_2^2) (-3 + 8 r - 24 r^2 + 24 r^3 - 5 r^4 - 12 r^2 \ln(r) ) 
 \nonumber
 \\
 &&\quad\quad\quad\quad\;\,
 + 2 C_1 C_2 (-19 + 56 r - 72 r^2 + 40 r^3 - 5 r^4 - 12 r^2 \ln(r)) \,,
 \\
  &&
  \nonumber
 \\
  &C_{\rho_D}^{{\scriptsize\overline{\mbox{MS}}}} &= 
  C_1^2 \bigg[-77 + 88 r - 24 r^2 + 8 r^3 + 5 r^4 - 48 \ln(r) - 36 r^2 \ln(r) \bigg] 
  \nonumber
 \\
 &&\quad
 + \frac{2}{3} C_1 C_2 \bigg[ -53 + 16 r + 144 r^2 - 112 r^3 + 5 r^4 + 96 (-1 + r)^3 \ln(1-r) 
 \nonumber
 \\
 &&\quad
 - 12 (4 - 9r^2 + 4r^3) \ln(r) 
 - 48 (-1 + r)^3 \ln\left(\frac{\mu^2}{m_b^2}\right) 
 \bigg]
 \nonumber
 \\
 &&\quad
 + C_2^2 \bigg[ -45 + 16 r + 72 r^2 - 48 r^3 + 5 r^4 + 96 (-1 + r)^2 (1 + r) \ln(1-r) 
 \nonumber
 \\
 &&\quad
 + 12 (1 - 4 r) r^2 \ln(r)
 - 48 (-1 + r)^2 (1 + r) \ln\left(\frac{\mu^2}{m_b^2}\right) \bigg]\,,
\end{eqnarray}
%
and for the case $\Gamma(b\rightarrow c\bar c s)$


%

\begin{eqnarray}
 &C_0& = C_{\mu_\pi} = (3 C_1^2 + 2 C_1 C_2 + 3 C_2^2) \bigg[ (1 - 14r - 2r^2 - 12 r^3 ) z 
 - 24 r^2 (-1 + r^2) \ln\left(\frac{1+z}{1-z}\right) \bigg]\,,
 \nonumber
 \\
 &&
 \\
 &C_v& = (3 C_1^2 + 2 C_1 C_2 + 3 C_2^2) \bigg[ 
 (5 - 38 r + 6 r^2 + 36 r^3) z 
 + 24 r^2(1 + 3 r^2)\ln\left(\frac{1+z}{1-z}\right)
   \bigg]\,,
 \\
 &&
 \nonumber
 \\
 &C_{\mu_G} &=  C_{\rho_{LS}} = -3(C_1^2 + C_2^2) \bigg[ (3 - 10 r + 10 r^2 + 60 r^3) z + 
    24 r^2 (-1 + 5 r^2) \ln\left(\frac{1+z}{1-z}\right) \bigg] 
  \nonumber 
  \\
 &&\quad\quad\quad\quad\;\,
 - 2 C_1 C_2 \bigg[ (19 - 2 r + 58 r^2 + 60 r^3) z 
 + 24 r (-2 - r + 4 r^2 + 5 r^3 )\ln\left(\frac{1+z}{1-z}\right) \bigg]\,,
 \nonumber
 \\
 &&
 \\
 &C_{\rho_D}^{{\scriptsize\overline{\mbox{MS}}}} &=
  C_1^2 \bigg[ (-77 - 2 r + 58 r^2 + 60 r^3) z + 
    24 (2 - 2 r - r^2 + 4 r^3 + 5 r^4) \ln\left(\frac{1+z}{1-z}\right) \bigg]
 \nonumber
 \\
 &&\quad
 + C_2^2 \bigg[ 24 (-4 + 8 r + 7 r^2 + 8 r^3 + 5 r^4) \ln\left(\frac{1+z}{1-z}\right)
  \nonumber
 \\
 &&\quad
 +  z \bigg( -45 - 58 r + 106 r^2 + 60 r^3 - 96 \ln(r) +  192 \ln(z) 
 - 48 \ln\left(\frac{\mu^2}{m_b^2}\right)  \bigg)\bigg] 
 \nonumber
 \\
 && \quad      
       + \frac{2}{3} C_1 C_2 \bigg[ 24 (-6 + 10 r - 5 r^2 + 20 r^3 + 5 r^4) \ln\left(\frac{1+z}{1-z}\right)
 \nonumber
 \\
&&\quad
    + z \bigg(75 - 178 r + 250 r^2 + 60 r^3  - 96 \ln(r) 
    + 192 \ln(z)  
    - 48 \ln\left(\frac{\mu^2}{m_b^2}\right)  \bigg) \bigg]\,.
\end{eqnarray}
%
%
%
%
%
Here $r=m_c^2/m_b^2$ and $z=\sqrt{1-4r}$. We note that the 
equalities $C_0 = C_{\mu_\pi}$ and $C_{\mu_G} =  C_{\rho_{LS}}$ are a consequence of reparametrization 
invariance~\cite{Mannel:2018mqv}.  
We take this here as a check of our calculation.  
We also note that the results for the coefficient of $\rho_D$ depend on 
the calculational scheme. 
This does not only concern the use of the $\overline{\rm MS}$ scheme, but also the treatment 
of the Dirac algebra in $D$ dimensions. This is related to the fact that the Fierz-rearrangement 
in $D$ dimensions generates 
evanescent operators which result in constants to be taken into account 
when comparing results~\cite{Lenz:2020oce}.  

After using the transformation rules (3.20-3.23) in~\cite{Turczyk:07}, and after proper definition of evanescent operators for the 
$b \rightarrow cud$ channel (see Sec.~\ref{Sec:eva}), these results are in agreement with~\cite{Lenz:2020oce}, where the coefficients were computed in 
four dimensions. We will comment in more detail on this in the next section. 

Note that from these results one readily finds the coefficients of HQE
in eq.~ (\ref{hqeTOPm3}) whose computation was described in Sec.~\ref{Sec:match2F}
\begin{eqnarray}
 C_\pi &=& C_{\mu_\pi} - C_0 \tilde C_{\pi} + C_v\,,
\\
 C_G &=&  C_{\mu_G} + C_0 \tilde C_G + C_v\,,
\\
 C_D^{{\scriptsize\overline{\mbox{MS}}}} &=& C_{\rho_D}^{{\scriptsize\overline{\mbox{MS}}}} + C_0 \tilde C_D + \frac{1}{2}C_v\,,
\\
 C_{LS} &=&  C_{\rho_{LS}} + C_0 \tilde C_{LS} + \frac{1}{2}C_v\,.
\end{eqnarray}

\subsection{Comment on the basis of four-quark operators}
\label{Sec:eva}
Our results discussed above are expressed in the operator basis
\begin{eqnarray}
 \mathcal{O}_{4F_1}^{(u)} &=& 
(\bar h_v \Gamma^\sigma \gamma^\mu \Gamma^\rho u) (\bar u \Gamma_\sigma \gamma_\mu \Gamma_\rho h_v)
 = (\bar h_v \gamma^\sigma \gamma^\mu \gamma^\rho P_L u) (\bar u P_R \gamma_\sigma  \gamma_\mu \gamma_\rho h_v)\,,
\\ 
 \mathcal{O}_{4F_2}^{(u)} &=& 
(\bar h_v \Gamma^\sigma \slashed v \Gamma^\rho u) (\bar u \Gamma_\sigma \slashed v \Gamma_\rho h_v)\\
 &=& (\bar h_v \gamma^\sigma \gamma^\rho P_L u) (\bar u P_R \gamma_\sigma  \gamma_\rho h_v)
+ 4 (\bar h_v \gamma^\rho P_L u) (\bar u P_R \gamma_\rho h_v)
- 4 (\bar h_v P_L u) (\bar u P_R h_v)\,, \nonumber
\end{eqnarray}
while one may chose as well the basis
\begin{eqnarray}
\tilde{\mathcal{O}}_{4F_1}^{(u)} &=& (\bar h_v \Gamma_\mu u) (\bar u \Gamma^\mu h_v)\,, \\
\tilde{\mathcal{O}}_{4F_2}^{(u)} &=& (\bar h_v P_L u) (\bar u P_R h_v)\,,
\end{eqnarray}
which has been used in ref.~\cite{Lenz:2020oce}. 
While the two bases are 
equivalent in $D = 4$, the situation for arbitrary $D$ is more 
involved. Relating the two bases in $D$ dimensions requires the addition of new operators called evanescent operators. The choice of the evanescent operator is not unique, and a particular recipe reduces to a 
substitution~\cite{Beneke:2002rj,Grozin:2017uto}
\begin{eqnarray}
&&  \gamma_\mu\gamma_\nu\gamma_\alpha P_L \otimes \gamma^\mu \gamma^\nu \gamma^\alpha P_L \rightarrow 
 (16-a\epsilon)\gamma_\alpha P_L \otimes\gamma^\alpha P_L + E_1^{QCD}\,, \\ 
&&  \gamma_\mu\gamma_\nu P_L \otimes \gamma^\mu \gamma^\nu P_R \rightarrow 
 (4-b\epsilon)P_L \otimes P_R + E_2^{QCD}\,.
\end{eqnarray}
A conventional choice is $a=4$ and $b=-4$, with $d=4-2\epsilon$. We will call the basis fixed by this choice 
to be the {\it canonical} basis of four-quark operators. The evanescent 
operators are thus defined as 
\begin{eqnarray}
 E_1^{QCD} &=& \gamma_\mu\gamma_\nu\gamma_\alpha P_L \otimes \gamma^\mu \gamma^\nu \gamma^\alpha P_L
 - (16-a\epsilon)\gamma_\alpha P_L \otimes\gamma^\alpha P_L\,, \\ 
 E_2^{QCD} &=& \gamma_\mu\gamma_\nu P_L \otimes \gamma^\mu \gamma^\nu P_R - (4-b\epsilon)P_L \otimes P_R\,.
\end{eqnarray}
The choice of the evanescent operators $E_{1,2}^{QCD}$ is not unique. 
This choice is motivated by the requirement of validity of Fierz transformation at one-loop order~\cite{Buras:1989xd,Beneke:2002rj}.

Thus the complete operator basis reads
\begin{eqnarray}
\tilde{\mathcal{O}}_{4F_1}^{(u)} &=& (\bar h_v \Gamma_\mu u) (\bar u \Gamma^\mu h_v)\,, \\ 
\tilde{\mathcal{O}}_{4F_2}^{(u)} &=& (\bar h_v P_L u) (\bar u P_R h_v)\,, \\ 
 E_1^{QCD} &=& (\bar h_v \gamma_\mu\gamma_\nu\gamma_\alpha P_L u) (\bar u \gamma^\mu \gamma^\nu \gamma^\alpha P_L h_v)
 - (16-a\epsilon)(\bar h_v \Gamma_\mu u) (\bar u \Gamma^\mu h_v)\,, \\ 
 E_2^{QCD} &=& (\bar h_v \gamma_\mu\gamma_\nu P_L u) (\bar u P_R \gamma^\mu \gamma^\nu h_v) 
 - (4-b\epsilon)(\bar h_v P_L u)(\bar u P_R h_v)\,,
\end{eqnarray}
and the rule for the transformation between the two bases is
\begin{eqnarray}
 \mathcal{O}_{4F_1}^{(u)} &=& (16-a\epsilon) \tilde{\mathcal{O}}_{4F_1}^{(u)} + E_1^{QCD}\,, \\ 
 \mathcal{O}_{4F_2}^{(u)} &=& 
 4 \tilde{\mathcal{O}}_{4F_1}^{(u)}
 - b\epsilon\tilde{\mathcal{O}}_{4F_2}^{(u)}
+  E_2^{QCD}\,.
\end{eqnarray}
In the new basis the imaginary part of the transition operator becomes
\begin{equation}
 \mbox{Im} \mathcal{T}(b\rightarrow c\bar u d) = \Gamma_{\bar q_1 q_2}^0
 \bigg(\ldots
 + \tilde{C}_{4F_1}^{(u)} \frac{\tilde{\mathcal{O}}_{4F_1}^{(u)}}{4m_b^3}  
 + \tilde{C}_{4F_2}^{(u)} \frac{\tilde{\mathcal{O}}_{4F_2}^{(u)}}{4m_b^3} 
 + C_{E_1}^{(u)} \frac{E_1^{QCD}}{4m_b^3}
 + C_{E_2}^{(u)} \frac{E_2^{QCD}}{4m_b^3}
 \bigg)\,,
\end{equation}
with
\begin{eqnarray}
 \tilde{C}_{4F_1}^{(u)} &=&  (16-a\epsilon) C_{4F_1}^{(u)} + 4C_{4F_2}^{(u)} \,,
\\
 \tilde{C}_{4F_2}^{(u)} &=& - b\epsilon C_{4F_2}^{(u)}\,,
\\
 C_{E_1}^{(u)} &=& C_{4F_1}^{(u)}\,,
\\
 C_{E_2}^{(u)} &=& C_{4F_2}^{(u)}\,.
\end{eqnarray}
The operators $E_{1,2}^{QCD}$ do not contribute to the anomalous dimension of 
$C_{\rho_D}$. However, the change of basis produces a shift in the $\rho_D$ coefficient which depends on the choice 
of the evanescent operators i.e. on $a$ and $b$. We call the new coefficient $C_{\rho_D}^{{\scriptsize\overline{\mbox{MS}}}}(a,b)$. 
The difference between the results obtained in the 
two bases is
\begin{equation}
 C_{\rho_D}^{{\scriptsize\overline{\mbox{MS}}}}(a,b) - C_{\rho_D}^{{\scriptsize\overline{\mbox{MS}}}}
 = \frac{8}{3} C_1 C_2 (1 - r)^2 ( a (1 - r) - b (1 + 2 r) )
\end{equation}
whereas the difference between our results and the ones obtained in ref. \cite{Lenz:2020oce}, 
where the coefficients are computed in $D=4$, is

\begin{equation}
 C_{\rho_D}^{{\scriptsize\overline{\mbox{MS}}}}(a,b) - C_{\rho_D}^{{\scriptsize\overline{\mbox{MS}}},\,D=4} 
 = \frac{8}{3} C_1 C_2 (1 - r)^2 ( 4(2 + r) + b(1 + 2r) - a (1 - r)  )\,.
\end{equation}
Note that for the canonical choice of the evanescent operators the difference vanishes. 

As we mentioned there is some freedom when choosing the evanescent operators $E_{1,2}^{QCD}$. 
The difference in the results due to the different choice of $a$ and $b$ corresponds to   
a shift in the coefficient 
\begin{equation}
 C_{\rho_D}^{{\scriptsize\overline{\mbox{MS}}}}(a_1,b_1) - C_{\rho_D}^{{\scriptsize\overline{\mbox{MS}}}}(a_2,b_2) 
 = -\frac{8}{3} C_1 C_2 (1 - r)^2 ( - (a_1-a_2) (1 - r) + (b_1 - b_2) (1 + 2 r))\,.
\end{equation}
When inserting numbers one has to keep in mind, that the coefficient is thus dependent on the scheme. 
This scheme dependence is compensated by the value of $\rho_D$ itself, which is a scheme-dependent 
quantity.   

\subsection{Numerical analysis}
In this section we give numerical values for phenomenological applications. We will chose the 
canonical scheme for the evanescent operators such that the results in~\cite{Lenz:2020oce} can be directly 
compared to our results. We employ the $\overline{\rm MS}$ scheme for the definition of $\rho_D$ 
and chose for the scale $\mu = m_b$. 

For both channels we have contributions which come from the operators ${\cal O}_1$ and ${\cal O}_2$, 
which come with the Wilson coefficients $C_1$ and $C_2$, see (~\ref{Leff}). In table~\ref{tabbccs} we give the 
numerical values of the coefficients for the transition $b \rightarrow c\bar c s$. We also
list the values of the coefficients for the transition $b \rightarrow c\bar u d$ in the four-quark operator basis of 
Sec. ~\ref{SubSec:match4qbcud} in table~\ref{tabbcud}, and in the canonical basis in table~\ref{tabbcudcan}.
\begin{table}[!ht]
 \centering
  \begin{tabular}{|c|c|c|c|c|}
    \hline
    $b \rightarrow c\bar c s$ & $C_1^2$ & $C_2^2$ & $C_1 C_2$  \\ \hline
    $C_0$ & $0.84$ & $0.84$ & $0.56$ \\ \hline
    $C_{\mu_\pi}$ & $0.84$ & $0.84$ &  $0.56$ \\ \hline
    $C_{\mu_G}$ & $-4.99$ & $- 4.99$ & $- 14.6$  \\ \hline
    $C_{\rho_D}$ & $44.0$ & $- 56.1$ &  $- 49.5$ \\ \hline
    $C_{\rho_{LS}}$ & $-4.99$ & $- 4.99$ & $- 14.6$  \\ \hline
    $C_{4F_1}^{(s)}/(128\pi^2)$ & $0$ & $- 9.35$ & $-6.23$ \\ \hline
    $C_{4F_2}^{(s)}/(128\pi^2)$ & $0$ & $11.6$ & $7.71$  \\ \hline
    \end{tabular}
    \caption{Numerical values for the coefficients of $b \rightarrow c\bar c s$. For illustration we take the numerical values 
    $\mu= m_b=4.8$~GeV and $m_c=1.3$~GeV.}
    \label{tabbccs}
\end{table}

\begin{table}[!ht]
	\centering
	\begin{tabular}{|c|c|c|c|c|}
		\hline
		$b \rightarrow c\bar u d$ & $C_1^2$ & $C_2^2$ & $C_1 C_2$  \\ \hline
		$C_0$ & $1.75$ & $1.75$ & $1.17$ \\ \hline
		$C_{\mu_\pi}$ & $1.75$ & $1.75$ & $1.17$  \\ \hline
		$C_{\mu_G}$ & $-7.09$ & $- 7.09$ & $- 30.2$  \\ \hline
		$C_{\rho_D}$ & $55.2$ & $- 50.3$ &  $52.4$ \\ \hline
		$C_{\rho_{LS}}$ & $-7.09$ & $- 7.09$ & $- 30.2$  \\ \hline
		$C_{4F_1}^{(d)}/(128\pi^2)$ & $0$ & $- 10.7$  & $-7.12$ \\ \hline
		$C_{4F_2}^{(d)}/(128\pi^2)$ & $0$ & $11.8$  & $7.88$  \\ \hline
		$C_{4F_1}^{(u)}/(128\pi^2)$ & $0$ & $0$ & $0.80$  \\ \hline
		$C_{4F_2}^{(u)}/(128\pi^2)$ & $0$ & $0$ & $1.97$  \\ \hline
	\end{tabular}
	\caption{Numerical values for the coefficients of $b \rightarrow c\bar u d$. For illustration we take the numerical values 
		$\mu= m_b=4.8$~GeV and $m_c=1.3$~GeV.}
	\label{tabbcud}
\end{table}
\begin{table}[!ht]
 \centering
  \begin{tabular}{|c|c|c|c|c|}
    \hline
    $b \rightarrow c\bar u d$ & $C_1^2$ & $C_2^2$ & $C_1 C_2$  \\ \hline
    $C_0$ & $1.75$ & $1.75$ & $1.17$ \\ \hline
    $C_{\mu_\pi}$ & $1.75$ & $1.75$ & $1.17$  \\ \hline
    $C_{\mu_G}$ & $-7.09$ & $- 7.09$ & $- 30.2$  \\ \hline
    $C_{\rho_D}$ & $55.2$ & $- 50.3$ &  $71.4$ \\ \hline
    $C_{\rho_{LS}}$ & $-7.09$ & $- 7.09$ &  $- 30.2$ \\ \hline
    $C_{4F_1}^{(d)}/(128\pi^2)$ & $0$ & $- 10.7$ & $-7.12$ \\ \hline
    $C_{4F_2}^{(d)}/(128\pi^2)$ & $0$ & $11.8$ & $7.88$  \\ \hline
    $\tilde C_{4F_1}^{(u)}/(128\pi^2)$ & $0$ & $0$ &  $20.6$ \\ \hline
    $\tilde C_{4F_2}^{(u)}/(128\pi^2)$ & $0$ & $0$ &  $0$ \\ \hline
    $C_{E_1}^{(u)}/(128\pi^2)$ & $0$ & $0$ &  $0.80$ \\ \hline
    $C_{E_2}^{(u)}/(128\pi^2)$ & $0$ & $0$ &  $1.97$ \\ \hline
    \end{tabular}
    \caption{Numerical values for the coefficients of $b \rightarrow c\bar u d$ in the canonical basis ($a=4$, $b=-4$). For illustration 
    we take the numerical values $\mu= m_b=4.8$~GeV and $m_c=1.3$~GeV.}
    \label{tabbcudcan}
\end{table}
In order to get an idea about the size of the total contribution of $\rho_D$ to the non-leptonic
width we insert values for 
the Wilson coefficients $C_1 (m_b) = -1.121$ and $C_2(m_b)  = 0.275$ (note that $C_1 (M_W) = -1$ and $C_2 (M_W) = 0$ to leading logs). 

We denote $\langle\mathcal{O}_{4F_i}^{(q)}\rangle \equiv \langle B(p_B) \lvert\mathcal{O}_{4F_i}^{(q)}\lvert B(p_B)\rangle/(2M_B)$ and use the 
abbreviation $\Gamma_{\bar q_1 q_2}$ to refer to $\Gamma(b\rightarrow c\bar q_1 q_2)$. We obtain
\begin{eqnarray}
 \frac{\Gamma_{\bar c s}}{\Gamma_{\bar c s}^0} &=&
 0.94 
 - 0.47 \frac{\mu_\pi^2}{m_b^2} 
 - 1.07 \frac{\mu_G^2}{m_b^2}
 - 33.2 \frac{\rho_D^3}{m_b^3} 
 + 1.07 \frac{\rho_{LS}^3}{m_b^3}
 + 383 \frac{\langle\mathcal{O}_{4F_1}^{(s)}\rangle}{m_b^3}
 - 475 \frac{\langle\mathcal{O}_{4F_2}^{(s)}\rangle}{m_b^3}\,,
\end{eqnarray}
\begin{eqnarray}
\frac{\Gamma_{\bar u d}}{\Gamma_{\bar u d}^0} &=&
 1.98
  - 0.99 \frac{\mu_\pi^2}{m_b^2}
 - 0.07 \frac{\mu_G^2}{m_b^2}
 - 24.7 \frac{\rho_D^3}{m_b^3}
 + 0.07 \frac{\rho_{LS}^3}{m_b^3}
 + 438 \frac{\langle\mathcal{O}_{4F_1}^{(d)}\rangle}{m_b^3} 
 - 485 \frac{\langle\mathcal{O}_{4F_2}^{(d)}\rangle}{m_b^3} 
 \nonumber
 \\
 &&
 - 77.5  \frac{\langle\mathcal{O}_{4F_1}^{(u)}\rangle}{m_b^3} 
 - 192 \frac{\langle\mathcal{O}_{4F_2}^{(u)}\rangle}{m_b^3} \,,
\end{eqnarray}
\begin{eqnarray}
 \frac{\Gamma_{\bar u d}}{\Gamma_{\bar u d}^0}\bigg|_{{\scriptsize{\mbox{can. basis}}}} &=&
 1.98
 - 0.99 \frac{\mu_\pi^2}{m_b^2}
 - 0.07 \frac{\mu_G^2}{m_b^2}
 - 21.8 \frac{\rho_D^3}{m_b^3} 
 + 0.07  \frac{\rho_{LS}^3}{m_b^3}
 + 438 \frac{\langle\mathcal{O}_{4F_1}^{(d)}\rangle}{m_b^3}
 - 485 \frac{\langle\mathcal{O}_{4F_2}^{(d)}\rangle}{m_b^3}
 \nonumber
 \\
 &&
 - 2\cdot 10^3 \frac{\langle \tilde{\mathcal{O}}_{4F_1}^{(u)}\rangle}{m_b^3} 
 - 77.5 \frac{\langle E_1^{QCD}\rangle}{m_b^3} 
 - 192 \frac{\langle E_2^{QCD}\rangle}{m_b^3}\,,
\end{eqnarray}
Assuming that $\langle E_1^{QCD}\rangle/m_b^3=\langle E_1^{QCD}\rangle/m_b^3=0$ and 
$\rho_D^3/m_b^3 \sim \langle \mathcal{O}_{4F_i}^{(q)}\rangle/ m_b^3\sim \Lambda_{QCD}^3/m_b^3$, the Darwin coefficient gives a correction to the tree 
level values of the coefficients of the four-quark operator of $\sim 7\%$ for $b\rightarrow c\bar cs$ and of $\sim 1\%$ for $b\rightarrow c\bar ud$ 
in the canonical basis (we take the largest coefficient of the four-quark operators to compare).

\section{Discussion and conclusions}
\label{sect:discussion}
We have computed the contributions of the Darwin and the spin-orbit term appearing 
at order $1/m_Q^3$ in the HQE of the non-leptonic width. 
Although the coefficient of the spin-orbit term is fixed by reparametrization invariance, 
we have explicitly computed it as a check of our methods.

The most interesting part is the computation of the coefficient of $\rho_D$, since  
one has to take into account the mixing with the four-quark operators. The coefficient 
of the Darwin term turns out to be sizable which was also found in the semi-leptonic 
case. 

In fact, this may become relevant for lifetime differences. To be specific, we will 
consider the $SU(3)_{\rm Flavor}$ triplet of ground state $B$ hadrons
${\cal B}=(B_u,B_d,B_s)$. It has been noticed already very early~\cite{Neubert:1996we} that up 
to and including $1/m_b^2$ the operators appearing in ${\rm Im} \, {\cal T}$ are $SU(3)_{\rm flavour}$ 
singlets and hence a lifetime difference to this order can only emerge from the $SU(3)$ breaking 
coming from the states, meaning that $\mu_\pi$ and $\mu_G$ differ between the three $B$-meson 
ground states. In turn, assuming the $SU(3)$ flavour symmetry, 
no lifetime differences can be induced 
up to this order. Since the coefficients of the HQE parameters at order $1/m_b^2$ are small, 
the effect on lifetime differences due to the $SU(3)_{\rm flavour}$ breaking in $\mu_\pi$ and $\mu_G$ 
is very small. 

The situation changes at the order $1/m_b^3$ where four-quark operators appear, involving light quarks. Since 
the weak hamiltonian is sensitive to the light-quark flavor, the resulting four-quark operators have 
different matching coefficients. Analyzing the $SU(3)_{\rm flavour}$ structure of the four-quark 
operators, we may decompose them into a singlet and an octet contribution with respect to  
$SU(3)_{\rm Flavor}$ according to  
\begin{equation}
 \bar{h}_v \Gamma (\bar{q} {\mathbf 1}  q)^T \Gamma h_v\,, \qquad 
\bar{h}_v \Gamma (\bar{q} {\mathbf T}^a  q)^T \Gamma h_v\,,
\end{equation}
where $q = (u,d,s)$ is the light quark triplet and ${\mathbf 1}$ and ${\mathbf T}^a$ are the 
generators of  $U(3)_{\rm flavour}$. If we, in addition, use the equation of motion 
\begin{equation}
{\rm Tr} \{ \lambda^a  [(iD_\mu ) , [(iD^\mu),(iD^\nu)]] \} = g \sum_q \bar{q} \gamma^\nu \lambda^a q \,,
\end{equation}
(where $\lambda^a$ are the Gell-Mann matrices of $SU(3)_{\rm color}$) we can eliminate $\rho_D$ in favour of four-quark operators, contributing to the $SU(3)_{\rm Flavor}$
singlet part only. Obviously the mixing of $\rho_D$ can only happen with the $SU(3)_{\rm Flavor}$ 
singlet part of the four-quark operators.  

Overall, we thus find that we can re-write~(\ref{Res1}) as (schematically)    
\begin{eqnarray}
\Gamma(b\rightarrow c \bar q_1 q_2) &=& \Gamma_{\bar q_1 q_2}^0 
\bigg[   C_0 \left( 1 -
 \frac{\mu_\pi^2}{2m_b^2} \right)
+ \frac{C_{\mu_G}}{2m_b^2} \left(\mu_G^2 - \frac{1}{m_b} \rho_{LS}^3  \right)  \\ \nonumber
&& \qquad + \frac{1}{m_b^3} \left( \sum_i C_{T,s,i} T^{4q}_{i,{\rm singlet}} + 
  \sum_j C_{T,o,j} T^{4q}_{j,{\rm octet}} \right)
\bigg]\,,
\end{eqnarray}
where the sums run over the matrix elements of the four-quark operators. 

Clearly the octet part of the four-quark operators is  the main source for lifetime differences, 
which is present also if the states are exactly $SU(3)$ symmetric. However, the precision of the 
lifetime measurements has increased and thus also the $SU(3)$ 
breaking through the states needs to be taken into account, which means that also the matrix 
elements $T^{4q}_{i,{\rm singlet}}$ will contribute to lifetime differences. This effect 
may be important, since the complete calculation of these terms shows that the 
coefficients of these terms are large, even enhanced by phase space factors. 
However, a quantitative study of their impact on lifetime differences needs estimates 
of the  $SU(3)_{\rm flavour}$ breaking in the matrix elements $T^{4q}_{i,{\rm singlet}}$
which is beyond the scope of the present paper.

\subsection*{Acknowledgments}
AP and ThM thank Andrey Grozin for discussions in the early stage of this work. 
We thank Alexander Lenz, Maria Laura Piscopo, and Aleksey V. Rusov for communicating their 
results to us prior to publication. We thank Keri Vos for reporting us missprints in the preprint version.
   
This research was supported by the Deutsche Forschungsgemeinschaft 
(DFG, German Research Foundation) under grant  396021762 - TRR 257 
``Particle Physics Phenomenology after the Higgs Discovery''. 
%
  
%
\section*{Appendix\label{sec:appendix}}
Here we collect some technical results used for the computations.

\subsection*{The decay $b\rightarrow c\bar c s$}
The most complicated part of the computation technically is the decay into two heavy quarks -- $c$-quarks.

We define the most general two-loop integral that can appear as
\begin{eqnarray}
 &&J(n_1,n_2,n_3,n_4,n_5)
 \nonumber
 \\
 &&
 \equiv \int \frac{d^D q_1}{(2\pi)^D}\frac{d^D q_2}{(2\pi)^D}
 \frac{1}{(q_1^2)^{n_1} ((p+q_1-q_2)^2 - m_c^2)^{n_2} (q_2^2-m_c^2)^{n_3}((q_1 + p)^2)^{n_4}((q_2+p)^2)^{n_5}}
 \nonumber
 \\
 &&
  = \int \frac{d^D q_1}{(2\pi)^D}\frac{d^D q_2}{(2\pi)^D}\frac{1}{D_1^{n_1}D_2^{n_2}D_3^{n_3}D_4^{n_4}D_5^{n_5}}\,,
 \label{basiceqIBP}
\end{eqnarray}
where $+i0$ prescriptions are assumed in the propagators and $p^2=m_b^2$. 
Using the program LiteRed~\cite{Lee:2012cn,Lee:2013mka}, one finds that the amplitude for the 
relevant  
diagrams can be expressed as a combination of the following 
three master integrals
\begin{eqnarray}
 \label{mi1sunset2loop}
 &J(0,1,1,0,0) &= \int \frac{d^D q_1}{(2\pi)^D}\frac{d^D q_2}{(2\pi)^D}
 \frac{1}{((p+q_1-q_2)^2 - m_c^2) (q_2^2-m_c^2)}
 \nonumber
 \\
 &&
 = \int \frac{d^D q_1}{(2\pi)^D}\frac{1}{q_1^2 - m_c^2}
 \int \frac{d^D q_2}{(2\pi)^D} \frac{1}{q_2^2-m_c^2}\, ,
 \\
 &&\nonumber
 \\
 &J(1,1,1,0,0) &= \int \frac{d^D q_1}{(2\pi)^D}\frac{d^D q_2}{(2\pi)^D}
 \frac{1}{q_1^2 ((p+q_1-q_2)^2 - m_c^2) (q_2^2-m_c^2)}
 \nonumber
 \\
 &&= 
 \int \frac{d^D q_1}{(2\pi)^D} \frac{1}{(p-q_1)^2 - m_c^2}
 \int\frac{d^D q_2}{(2\pi)^D}
 \frac{1}{(q_1-q_2)^2  (q_2^2-m_c^2)}\, ,
 \\
 &&\nonumber
 \\
 &J(2,1,1,0,0) &= \int \frac{d^D q_1}{(2\pi)^D}\frac{d^D q_2}{(2\pi)^D}
 \frac{1}{q_1^4 ((p+q_1-q_2)^2 - m_c^2) (q_2^2-m_c^2)}
 \nonumber
 \\
 &&= 
 \int \frac{d^D q_1}{(2\pi)^D}
  \frac{1}{(p-q_1)^2 - m_c^2}
 \int\frac{d^D q_2}{(2\pi)^D}\frac{1}{(q_1-q_2)^4  (q_2^2-m_c^2)}\,.
  \label{mi3sunset2loop}
\end{eqnarray}
We are only interested in the imaginary part of the corresponding integrals, 
related to the discontinuity across the cut. We denote 
$\bar J \equiv \mbox{Im}\, J$. On the one hand $\bar J(0,1,1,0,0) = 0$. 
On the other hand, we can use that
\begin{equation}
 \frac{d}{dm_c^2} J(1,1,1,0,0) = J(1,2,1,0,0) + J(1,1,2,0,0) = 2J(1,2,1,0,0)\,,
\end{equation}
and the reduction of $J(1,2,1,0,0)$ to a combination of the three master integrals above
\begin{eqnarray}
 J(2,1,1,0,0) &=& - \frac{1}{(-4+D)m_b^2 (m_b^2 - 4m_c^2)}\bigg[
 (-2+D)^2 J(0,1,1,0,0) 
 \nonumber
 \\
 &&
 + (-3+D)(-8+3D)(m_b^2-2m_c^2)J(1,1,1,0,0) 
 \nonumber
 \\
 &&
 + 8(-3+D)m_c^2 (m_c^2 - m_b^2)J(1,2,1,0,0)
  \bigg]\,,
\end{eqnarray}
in order to express the master integral $J(2,1,1,0,0)$ only in terms of 
$J(1,1,1,0,0)$ and $J(0,1,1,0,0)$ whose imaginary part is zero
\begin{eqnarray}
 J(2,1,1,0,0) &=& - \frac{1}{(-4+D)m_b^2 (m_b^2 - 4m_c^2)}\bigg[
 (-2+D)^2 J(0,1,1,0,0) 
 \nonumber
 \\
 &&
 + (-3+D)(-8+3D)(m_b^2-2m_c^2)J(1,1,1,0,0) 
 \nonumber
 \\
 &&
 + 8(-3+D)m_c^2 (m_c^2 - m_b^2)\frac{1}{2}\frac{d}{dm_c^2}J(1,1,1,0,0)
  \bigg]\,.
  \label{m3m4}
\end{eqnarray}
Therefore there is only one master integral we need to compute, which is 
$J(1,1,1,0,0)$. Note that eq.~(\ref{m3m4}) has a pole in 
$D=4-2\epsilon$ dimensions. Therefore the $\mathcal{O}(\epsilon)$ expansion of 
$J(1,1,1,0,0)$ will be needed. 
We find
\begin{eqnarray}
 \bar J(1,1,1,0,0) &=& \frac{2^{-9+6\epsilon} 
\pi^{-3/2+2\epsilon} \csc(\pi\epsilon)}{\Gamma(3/2-\epsilon)}m_b^{2-4\epsilon} \bigg[
 \frac{ r^{1-\epsilon} (-5+2\epsilon)H_1}{\Gamma(3-\epsilon)}
 \nonumber
 \\
 &&
 + \frac{ r^{1-\epsilon} (1 + 4r(1-4\epsilon))H_2}{\Gamma(3-\epsilon)}
 \nonumber
 \\
 &&
 - \frac{ \Gamma(1-\epsilon)}{(-1+2\epsilon)\Gamma(3-3\epsilon)}\bigg(\frac{H_3}{\Gamma(-1+\epsilon)} + \frac{(2-3\epsilon)H_4}{\Gamma(\epsilon)}\bigg)
 \bigg]\,,
\end{eqnarray}
where
\begin{eqnarray}
 &H_1& = \,_2 F_1 (-1/2 , 2\epsilon, 3 - \epsilon, 4r )\,,
 \\
 \nonumber
 &H_2& =\,_2 F_1 (1/2, 2\epsilon, 3 - \epsilon, 4r)\,,
 \\
 \nonumber
 &H_3& =\,_2 F_1 ( -1/2 + \epsilon, -2 + 3 \epsilon, -1 + \epsilon, 4r )\,,
 \\
 \nonumber
 &H_4& =\,_2 F_1 ( -1/2 + \epsilon, -1 + 3\epsilon, \epsilon, 4r )\,.
\end{eqnarray}

\subsection*{The computation of the decay $b\rightarrow c\bar c s$ in a scheme with a hard IR regularization}
The case we are considering can be used for pedagogical purposes of OPE (HQE).
There is only an IR divergence in the original amplitude.
One can proceed by regulating IR with a small mass and use four dim to compute 
the coefficients. The relevant master integral is then defined
as follows
\begin{eqnarray}
J(0,1,1,0,2;m_0)|_{d=4} \equiv
\int
\frac{d^4 q_1 d^4 q_2}{ 
((p+q_1-q_2)^2 - m_c^2) (q_2^2 - m_c^2)
((q_2+p)^2-m_0^2)^2}\, .
\end{eqnarray}
Here $m_0$ is an IR regulator, $p^2=m_b^2$, and $m_0\ll m_b$. 
The HQE for the non-leptonic correlator at the order $1/m_b^3$
has the general form
\begin{equation}
\label{basiceqIBP1}
Amp = C_4{\cal O}_4 + C_D{\cal O}_D \, .
\end{equation}
For the sake of demonstration we take 
the only four-quark operator 
${\cal O}_4={\bar b} \gamma_\mu s {\bar s} \gamma_\mu b$.
The coefficient $C_4$ is 
computed in four dimensions and is finite in the limit $m_0/m_b\to 0$.
The expression for the $\rho_D$ coefficient $C_D$ is also obtained in four 
dimensions but contains
an IR singularity in the limit $m_0/m_b\to 0$. 

Thus the HQE becomes
\begin{equation}
\label{basiceqIBP1a}
Amp = C_4{\cal O}_4 +\left(C_D^{\rm finite} + C_4 \ln(m_0/m_b)\right){\cal O}_D
\, .
\end{equation}
Now one defines
the finite matrix element of 
the ${\bar b} \gamma_\mu s {\bar s} \gamma_\mu b$ operator as
(MS-subtracted)
\begin{equation}
\langle b|({\bar b} \gamma_\mu s {\bar s} \gamma_\mu b)^{\rm R}|gb\rangle \sim
\bigg(\frac{1}{\varepsilon}+\ln(m_0/\mu)+c\bigg)-\frac{1}{\varepsilon}\, .
\end{equation}
Upon substituting this expression in eq.~(\ref{basiceqIBP1})
one gets the finite coefficient $C_D^{\rm finite}$.
This is a rough sketch of the procedure used in~\cite{Lenz:2020oce}
(for a related discussion, see~\cite{Beneke:2002rj}).
In such an approach one needs an expansion of the master integral at 
the limit of small $m_0/m_b$.

We have obtained an analytical expression for the required expansion  
in the form
\begin{equation}
 \bigg[J(0,1,1,0,2;m_0)|_{d=4} - z\ln(m_0/m_b)\bigg]_{m_0/m_b\rightarrow 0} 
= 2(1-\rho)\ln\bigg(\frac{1+z}{1-z}\bigg) + z(1+2\ln(\rho)-4\ln (z) )\,,\nonumber
\end{equation}
where $r\equiv \rho = m_c^2/m_b^2$ and $z=\sqrt{1-4\rho}$.

%

\newpage

\end{document}